\documentclass{pazh7468}
\usepackage{graphicx}
\usepackage{latexsym}

\newcommand{\km}{\,\mbox{km}\,\mbox{s}^{-1}}
\def\Ha{\hbox{H$_\alpha$\,}}
\def\Hb{\hbox{H$_\beta$\,}}
\def\farcm{\hbox{$.\mkern-4mu^\prime$}}

\def\deg{\hbox{$^\circ$}}

\begin{document}

\title{NGC 7468: a galaxy with an inner polar disk}

\author{
L.V. Shalyapina$^1$ \and A.V. Moiseev$^2$ \and V.A.Yakovleva$^1$ \and
V.A. Hagen-Thorn$^1$  \and O.Yu. Barsunova$^1$}

\institute{ $^1$ Astronomical Institute, St. Petersburg State
University, Universitetsky pr.28, Petrodvorets, 198504 Russia
\\
$^2$ Special Astrophysical Observatory, Russian Academy of Sciences,
Nizhnii Arkhyz, Karachai-Cherkessian Republic, 357147 Russia }

\offprints{L.V.~Shalyapina, \email{lshal@astro.spbu.ru}}
\date{Received January 22, 2004}

\titlerunning{NGC 7468: a galaxy with an inner polar disk}
\authorrunning{ Shalyapina et al.}

\abstract{ We present our spectroscopic observations of the galaxy
NGC 7468 performed at the 6-m Special Astrophysical Observatory
telescope using the UAGS long-slit spectrograph, the  multipupil
fiber spectrograph MPFS, and the  scanning Fabry-Perot interferometer (IFP).
We found no significant  deviations from the circular rotation of the
galactic disk in the velocity field in the regions of brightness
excess along the major axis of the galaxy (the putative polar ring).
Thus, these features are either tidal structures or weakly developed
spiral arms. However, we detected a gaseous disk at the center of the
galaxy whose rotation plane is almost perpendicular to the plane of
the galactic disk. The central collision of NGC 7468 with a gas-rich
dwarf galaxy and their subsequent merging seem to be responsible for
the formation of this disk.} \maketitle

\centerline{\large \bf INTRODUCTION}

\medskip

 Whitmore et al. (1990) included NGC 7468 (Mrk 314) in
their catalog of polar-ring galaxies (PRGs), candidate PRGs, and
related objects as a probable candidate (C-69). The direct images
(Fig. 1) of this galaxy show an extended low surface brightness
base and a bright nuclear region resolvable into several
individual condensations. Lobes are
observed along the major axis on the galaxy's southern and
northern sides, which may suggest the existence of a ring; this
was the reason why NGC 7468 was included in the above catalog.
The southern protrusion transforms into a faint bar that ends
with a brightening whose distance from the galaxy's center in the
plane of the sky is $\sim45''$, or $\sim 7$ kpc. (For the
line-of-sight velocity we found, $V_{gal} = 2220\km$, and $H_0 =
65\km\,\mbox{Mpc}^{-1}$, the distance to the galaxy is 34 Mpc,
which yields a scale of 0.16 kpc in 1$''$).

NGC 7468 was initially classified as a peculiar elliptical galaxy
(RC3, LEDA). However, according to a detailed photometric study by
Evstigneeva (2000), it should be considered to be a late-type spiral
or an irregular galaxy. Indeed, the galaxy is rich in gas, as
follows from neutral and molecular hydrogen data (Taylor et al.
1993, 1994; Richter et al. 1994; Wiklind et al. 1995). The ratios
$M_{HI}/L_B = 0.621$ and $M_{H_{2}}/L_{B} = 0.06$ $(M_{H_{2}}/M_{HI}
= 0.11)$ derived for NGC 7468 are characteristic of galaxies of late
morphological types (Young et al. 1989). The galaxy is an infrared
source ($L_{IR}/L_B \sim 0.6$), suggesting the presence of dust whose
mass is estimated to be $0.4\cdot10^6M_\odot$ (Huchtmeier et al.
1995).

In the central part of the galaxy, Petrosyan et al. (1979)
distinguished three condensations whose emission spectra were
studied by Petrosyan (1981) and subsequently by Nordgren et al.
(1995). Since the spectra for the galaxy's central region are
similar to those for H II regions, beginning with the works by Thuan
et al. (1981), this object has been attributed to H II galaxies or
blue compact dwarf galaxies (BCDGs) with active star formation in
their central regions. The estimates of the star formation rate
(Evstigneeva 2000; Cairos et al. 2001b) correspond to the star
formation rate in BCDGs.

The velocity field inferred from neutral hydrogen (Taylor et al.
1993, 1994) exhibits no features in any of the two lobes and
in the brightening region. These authors pointed out that the
central isovelocities are distorted (oval distortion), which may be
indicative of the presence of a bar. However, the spatial resolution
of these observations is not high.

The photometry performed by Evstigneeva (2000) and Cairos et al.
(2001b) show that the color of the northern and southern lobes
and the brightening is bluer than that in the remaining regions of
the galaxy (except for the nuclear region), which was considered by
Evstigneeva (2000) as evidence for the presence of an outer polar
ring around NGC 7468.

\begin{figure}
\centerline{\includegraphics[width=7  cm]{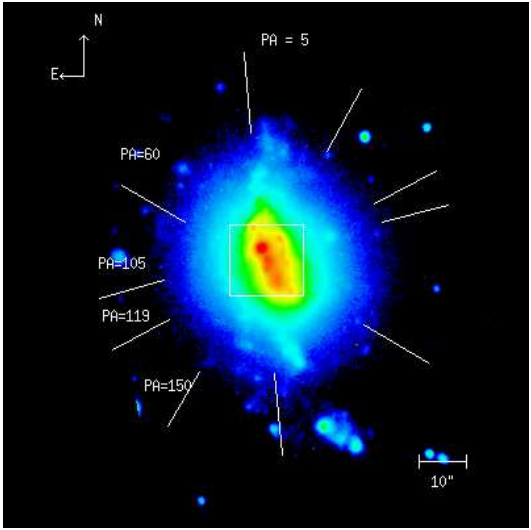}}
 \caption{Image of the galaxy NGC7468 obtained with the 6 m telescope through a filter centered
 at $\lambda6622$\AA; the UAGS slit locations and the MPFS field during
observations are shown.}
\end{figure}

 The initial assumption that the brightening to the south of the galaxy is a
superassociation (Petrosyan 1981) was replaced by Evstigneeva
(2000) with the suggestion that this is a companion galaxy,
probably because galaxy interaction is believed to be the most
probable cause of the polar-ring formation and because she failed
to find another suitable candidate that would confirm the
interaction. Meanwhile, Taylor et al. (1993) pointed out that
there is a faint galaxy 4\farcm6 (45.5 kpc) north of NGC 7468
that the authors believe to be a possible companion of NGC 7468.
The currently available information is clearly insufficient to
determine whether NGC 7468 belongs to PRGs. The velocity field of
the galaxy should be studied with a high spatial resolution. This
paper presents the results of such a study based on optical
spectroscopy.

\bigskip

\centerline{\large \bf OBSERVATIONS AND DATA REDUCTION}

\medskip

We performed our spectroscopic observations of the galaxy NGC 7468
at the prime focus of the 6 -m Special Astrophysical Observatory
(SAO) telescope with the long-slit spectrograph (UAGS), the
multipupil fiber spectrograph (MPFS) (Afanasiev et al. 2001), and
the scanning Fabry-Perot interferometer (IFP) (see Moiseev (2002)
and the SAO
WWW-site\footnote{http://www.sao.ru/hq/lsfvo/devicese.html}). The
table 1 gives a log of observations.

The reduction technique that we used was described previously
(Shalyapina et al. (2004).

The long-slit spectra were taken near the \Ha line (see the table
1). The \Ha and [SII] $\lambda\lambda6716, 6730$\AA\, emission
lines proved to be brightest in the spectra; in addition, the
[NII]$\lambda\lambda6548, 6583$ \AA, HeI $\lambda6678 \AA$ lines
were observed. The line-of-sight velocity curves were constructed
from all lines. The accuracy of these measurements is $\pm10\km$.

The central region of the galaxy was observed with the MPFS in the
green spectral range. This range includes metal absorption lines
(MgI$\lambda5175$ \AA, FeI$\lambda5229\AA$, FeI+CaI
$\lambda5270\AA$, etc.) and the \Hb and
[OIII]$\lambda\lambda4959,5007$\AA\, emission lines. Spectra from
240 spatial elements that form a $16\times15$ array in the plane
of the sky were taken simultaneously.

Based on emission lines, we constructed the two dimensional maps of
the intensity and line-of-sight velocity distributions (velocity
fields). The line-of-sight velocities were determinations with
accuracy of $10-15\km$. We used the spectra of the twilight sky and
the galactic nucleus as templates for cross-correlation when
constructing the velocity field for the stellar component. The
errors in the line-of-sight velocities determined from absorption
lines were found to be $\sim15-20 \km$.

Our IFP observations were performed near the \Ha line.
Premonochromatization  was made using a filter with a central
wavelength of $6578$ \AA\, and $FWHM=19$\AA. The readout was made
in $2\times2$-pixel binning mode, with $512\times512$-pixel images
being obtained in each spectral channel. These data were used to
construct the velocity field and the \Ha and continuum brightness
maps. The measurement errors of the line-of-sight velocities do
not exceed $10 \km$.

\begin{table*}
\begin{center}
\caption {Log of observations of NGC 7468} \vskip 0.2cm
\begin{tabular}{rrcccrrr}     \hline \hline
Instrument, date&Exposure time, &Field & Scale, & Spectral & Seeing& Spectral& P.A. \\
 date&  s& $''$& $''$/px& resolution, \AA& $''$&  region, \AA& field\\
\hline
 UAGS        & 1800     &   2x140   &0.4& 3.6&1.3      &6200-7000   &$5^{\circ}$ \\
             & 1200     &   2x140   &0.4& 3.6&1.5      &6200-7000   &$60^{\circ}$ \\
2-4.10.99    & 1200     &   2x140   &0.4& 3.6&1.5      &6200-7000   &$105^{\circ}$ \\
             & 1800     &   2x140   &0.4& 3.6&1.3      &6200-7000   &$150^{\circ}$ \\
             & 300      &   2x140   &0.4& 3.6&1.3      &6200-7000   &$119^{\circ}$ \\
\hline
  MPFS       & 1800     &  16x15    & 1 & 4.5&2.0      &4600-6000   & center\\
 08.12.01    &          &           &   &    &        &             &\\
\hline
IFP          &$32\times180$&$5'\times5'$& 0.56& 2.5& 1.3   &$H_\alpha$  &\\
05.09.2002   &             &            &     &    &     &            &\\
\hline \hline
\end{tabular}
\end{center}
\end{table*}

\bigskip

\begin{center}
\large \bf RESULTS OF THE UAGS OBSERVATIONS AND THEIR ANALYSIS
\end{center}

\medskip

 The line-of-sight velocity curves constructed from different emission
lines are similar. Figure 2 shows the data obtained from the \Ha
line at five locations of the UAGS slit. The point of maximum
brightness that coincides with the southernmost of the three
condensations was taken as zero. As we see from the figures, all
line-of-sight velocity curves are peculiar. Let us consider these
peculiarities in more detail.

The line-of-sight velocity curve obtained at a slit location close
to the galaxy's major axis ($PA=5^\circ$) is shown in Fig. 2a. In
general, its shape agrees with that expected for the rotation of the
galaxy's gaseous disk around its minor axis. In the region
$-18''\leq R\leq 20''$, the line-of-sight velocities change from
$2140 \km$ to $2030 \km$, flattening out at $R \geq 20''$. However,
the change in the velocity is not monotonic. The waves that probably
associated with individual HII regions are seen in the curve at
$R\approx-3''-5''$ and $R\approx12''$, and a step where the velocity
is almost constant is noticeable at $0\leq R \leq 7''$. Obviously,
the gas responsible for the emission here is not involved in the
rotation of the galaxy's main disk.

The presence of a kinematically decoupled gaseous subsystem at the
center is confirmed by the line-of- sight velocity curves for
position angles close to the minor axis ($PA = 105^\circ$,
$PA=119^\circ$). In both cases, a rectilinear segment with a
gradient $dV/dR\approx 200 \km\,\mbox{kpc}^{-1}$  is clearly seen
in the region $0\leq R \leq 3''$, suggesting the rigid-body
rotation of the gas around the galaxy's major axis. To the east
of the center, the line-of-sight velocities are almost constant
and, on average, equal to $2115 \km$. The emission here is
attributable to the gas of the galaxy's disk that rotates around
its minor axis. At $R \geq3''$, the behavior of the line-of-sight
velocities is different on these two cuts. At the position angle
$PA = 105^\circ$, the slit apparently goes outside the region
where the emission of the gas belonging to the central subsystem
dominates, and the line-of-sight velocity begins to increase,
approaching  $2100 \km$, the value that the subsystem of the
galaxy's gaseous disk must have here. At $PA= 119^\circ$, the
further decrease in the line-of-sight velocity with a smaller
gradient ($30 \km\,\mbox{kpc}^{-1}$) probably stems from the fact
that the slit does not pass through the dynamic center of the
decoupled subsystem.

The line-of-sight velocity curves at $PA= 60^\circ$ and $PA=
150^\circ$  are consistent with the assumption that a gaseous
subsystem rotating around the major axis is present in the central
part of the galaxy. At $R \leq 3''$, the curve also has a
rectilinear segment with a gradient $dV/dR \approx 78 \km\,
\mbox{kpc}^{-1}$ northeast of the center at $PA= 60^\circ$ and with a
gradient $dV/dR\approx80 \km\mbox{kpc}^{-1}$ at $PA= 150^\circ$. The
outer segments of the line-of-sight velocity curves are determined
by the galaxy's gaseous disk.

\begin{figure*}
\centerline{\includegraphics[width=14  cm]{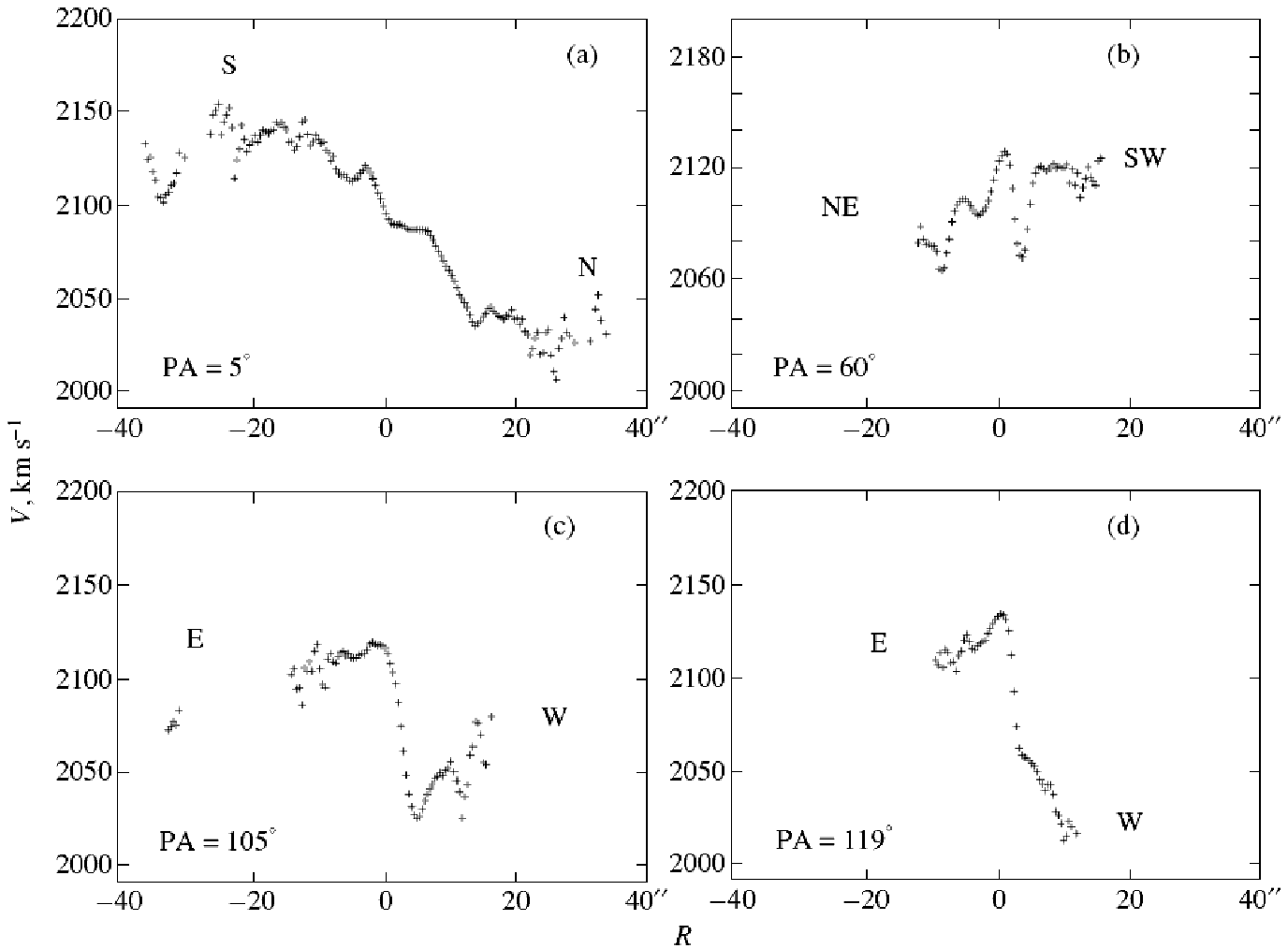}}
\centerline{\includegraphics[width=7  cm]{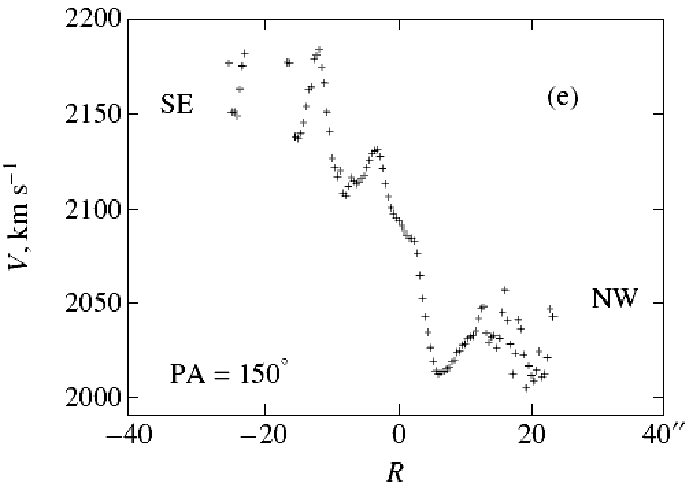}}
 \caption{NGC 7468: the UAGS line-of-sight velocity curves.}
\end{figure*}

 Thus, the behavior of all line-of-sight velocity curves indicates that there is a kinematically
 decoupled gaseous subsystem rotating around the major axis in the nuclear
region of the galaxy. The change in the line-of-sight velocity
gradient yields a preliminary estimate for the location of the
dynamic axis of this gaseous subsystem ($PA\approx110^\circ$). Its
size is about $6''$ (1 kpc). At $R \geq 10''$, the line-of-sight
velocity curves demonstrate the rotation of the galaxy's gaseous
disk around its minor axis; the locations of the photometric axis on
the periphery and the kinematic axis coincide and are close to $PA=
5^\circ$. An improvement of the kinematic features for the two
gaseous subsystems based on 2D spectroscopy will be described below.

Our comparative analysis of the line intensities shows that the line
intensity ratio $I_{[NII]\lambda6583}/I_{H\alpha}$ does not exceed
0.4 everywhere, whence it follows that the emission is produced by
photoionization. In this case, we can use the empirical method
described by Denicolo et al. (2001) to determine the metallicity:
\begin{equation}
  \begin{array}{c}
 12 + \log (O/H) = 9.12(\pm 0.05) + 0.73(\pm0.10) \times \\
\log(([SII]\lambda6716 + [SII]\lambda6731)/\mbox{H}_\alpha)\\
  \end{array}
\end{equation}

As an example, Fig~3 shows the radial distributions of $12 +
\log(O/H)$ for two cuts.  An inner region ($r \leq 5''$) where this
parameter is equal to $\sim8.6$ and $\sim8.4$ is distinguished in all
distribution; farther out, it increases and reaches $\sim 8.9$ in
outer regions. These values correspond to metallicities of
$0.5Z_\odot$, $0.3Z_\odot$, and $0.9Z_\odot$, respectively. The
spectrum along the galaxy's major axis ($PA= 5^\circ$) passes
through the brightest condensations designated by Petrosyan (1981)
as ``b'' and ``c''. In these regions, $12 + \log(O/H)\approx 8.6\pm
0.1$ ($\sim 0.5Z_\odot$). The distribution constructed for $PA=
119^\circ$ (Fig. 3b) exhibits the lowest values, $12 + \log(O/H) =
8.4$ ($\sim0.3Z_\odot$), at distances $R \leq 4''$ from the center.
These region are located in the area of the kinematically decoupled
gaseous subsystem, and the data for them apparently suggest that the
metallicity of the central subsystem is slightly lower.

\begin{figure*}
\centerline{\includegraphics[width=14  cm]{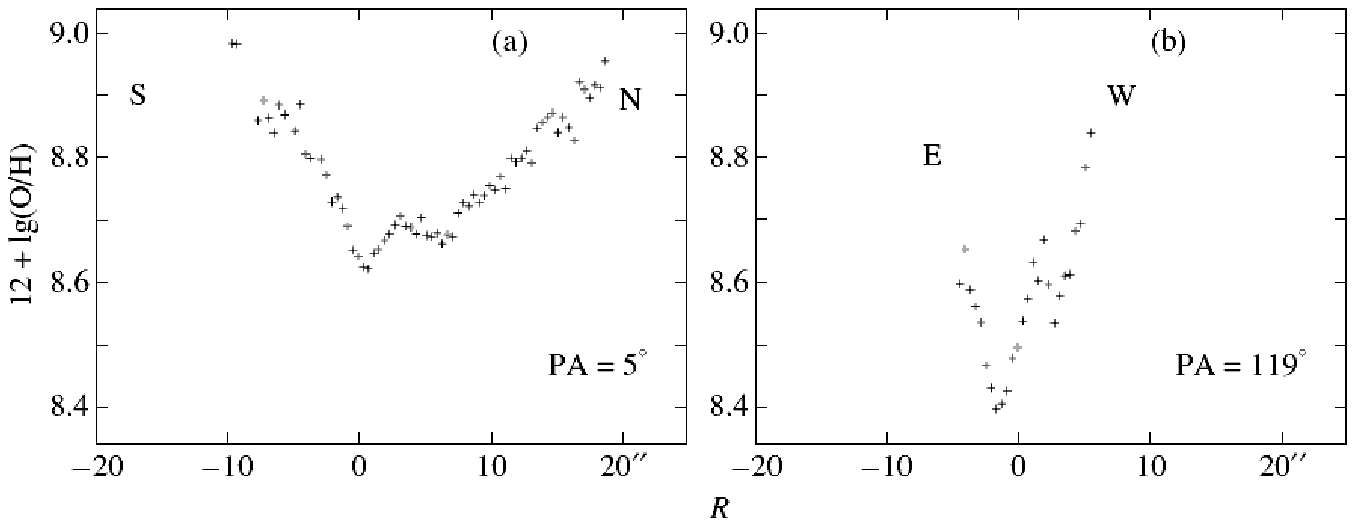}}
 \caption{Radial metallicity distributions.}
\end{figure*}

\bigskip

\begin{center}
\large \bf THE MORPHOLOGY OF NGC 7468
\end{center}

\medskip

The IFP and MPFS intensity maps in the emission lines and in the
continuum are shown in Fig. 4. The overall pattern of the brightness
distributions in the red continuum and in the \Ha line (Figs. 4a and
4b) demonstrates similarities, but numerous condensations of
different sizes, which are probably HII regions, are distinguished
more clearly in the \Ha image. Chains of HII regions are observed on
the southern side of the galaxy, and the southern lobe turns
toward the brightening, where three bright condensations are
distinguished. A lobe in the \Ha brightness distribution and
individual HII regions are also observed on the opposite side of the
galaxy. Note that the HII regions in the central part of NGC 7468
form a ring of a roughly oval shape. This structure also shows up in
the continuum brightness distribution, creating the illusion of a
barlike structure (Taylor et al. 1994).

The galaxy's outer amorphous base has roughly elliptical isophotes
with close axial ratios in the continuum ($b/a\approx0.7$) and in the
\Ha line ($b/a\approx0.65$); the geometrical center of the ellipses
is $3''$ west of condensation ``a''. The lobes directed
northward and southward are clearly seen in the continuum brightness
distribution.

The three bright knots that correspond to condensations ``a'', ``b'',
and ``c'' noted by Petrosyan et al. (1978) are clearly seen in the
central part of the image in the green continuum (Fig.
4c).Condensation ``c'' is brightest. In the emission lines (Fig. 4),
condensation ``c'' is poorly seen, while condensation ``a'' is most
prominent. This characteristic feature was noted by Petrosyan (1981).

\begin{figure*}
\centerline{\includegraphics[width=14  cm]{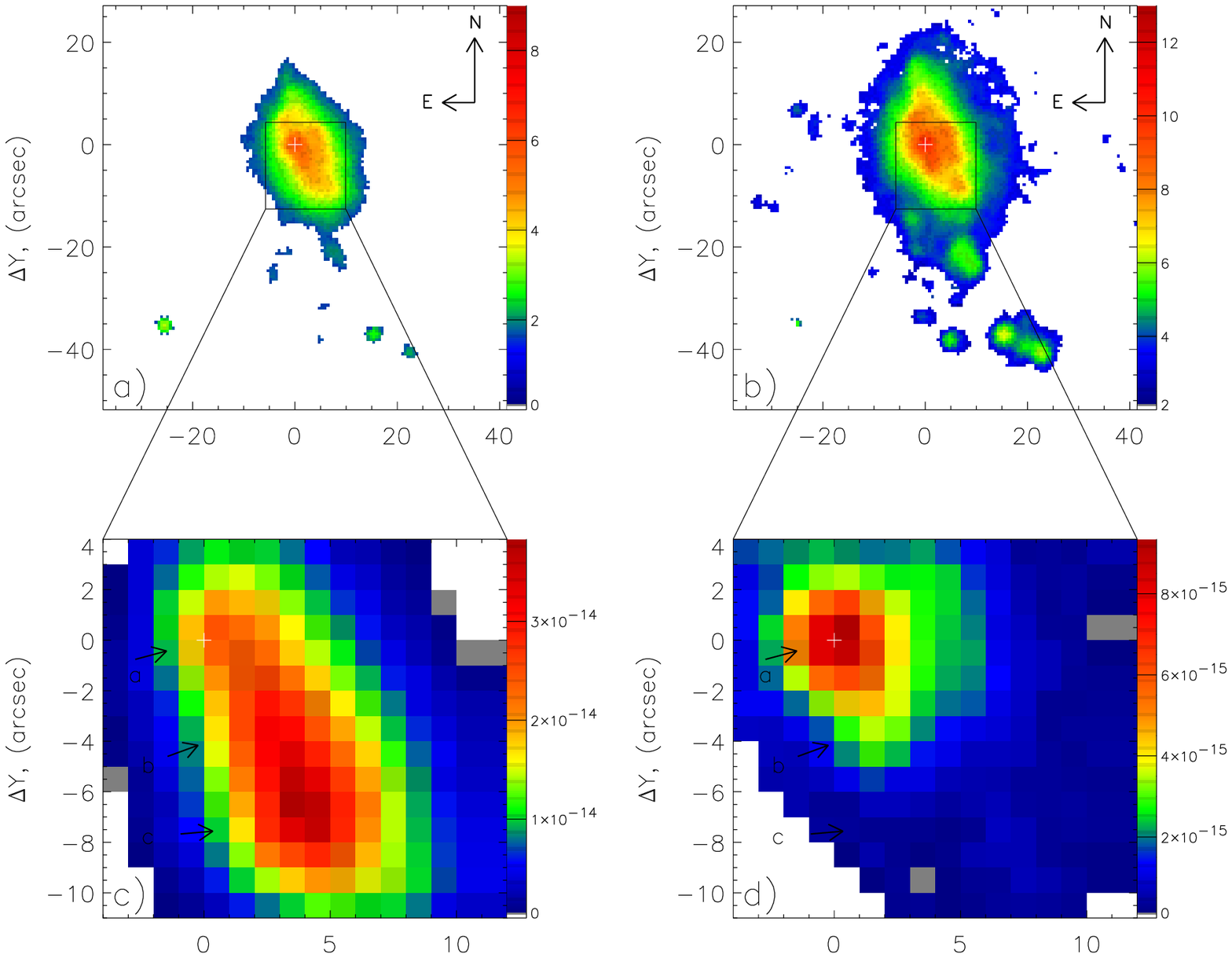}}
 \caption{IFP data: (a) images in the continuum near \Ha;
(b) an image in \Ha. MPFS data: the brightness distribution (c) in
the continuum ($\lambda5400--5500$\AA) and (d) in the
[OIII]$\lambda5007$ line. The gray scale corresponds to intensities
in arbitrary units  for panels (a) and (b) and to intensities in
$erg\, s^{-1}\, cm^2\, arcsec^{-2}\,$\AA$^{-1}$ for panels (c) and
(d). The pluses mark the position of the brightness maximum in the
red continuum.}
\end{figure*}

\begin{center}
\large \bf THE VELOCITY FIELDS OF THE IONIZED GAS AND STARS
\end{center}

Figure 5a shows the large-scale velocity field for the emitting gas,
as constructed from the IFP data, suggesting the rotation of the
galaxy's gaseous disk around its minor axis. However, a feature
indicating the rotation of the gas around the galaxy's major axis is
clearly seen in the central part of the velocity field. Thus, the IFP
data confirm the presence of two kinematic subsystems of gas in this
galaxy.

If the gas is assumed to rotate in circular orbits, then the method
of ``tilted-rings'' can be used (Begeman 1989; Moiseev and Mustsevoi
2000) to analyze the velocity field. The method allows us to
determine the positions of the kinematic center and the kinematic
axis (the direction of the maximum velocity gradient), to estimate
the galaxy's inclination, and to construct the rotation curve. As we
see from Fig. 6b, the location of the kinematic axis changes with
increasing radius: $PA_{dyn} \approx120^\circ$ in the nuclear region
($R \leq 6''$) and $PA_{dyn}\approx 180^\circ$ on the periphery. This
confirms the existence of two kinematic subsystems of gas. The
locations of the dynamic axes of the subsystems are close to those
estimated from long-slit spectra. The subsystem associated with the
central region is apparently the inner disk (ring) rotating around
the galaxy's major axis. The dynamic centers of the subsystems
coincide, but are displaced by approximately $3''$ to the W of
condensation ``a''. The heliocentric line-of-sight velocity of the
galaxy is $V_{sys} = 2070 \km$ or, after being corrected for the
rotation of our Galaxy, $V_{gal} = 2220 \km$. The observed velocity
field is best described by the model of  circular rotation at an
inclination of the galactic disk $i_{disk}\approx 45^\circ$ (which
is equal to the estimate obtained from the galaxy's axial ratio) and
that of the inner disk/ring, $i_{ring}\approx60''$. Here, two values
of the angle between the galactic disk and the plane of the inner
disk are possible: $\Delta i\approx 50^\circ$ and $\Delta i\approx
86^\circ$. In the first case, the orbits of the gas are unstable,
and, therefore, noncircular motions must be observed in the velocity
field. However, such motions have not been found, so the second
value (86\deg) of the angle between the galactic disk and the plane
of the inner disk should be taken.

\begin{figure*}
\centerline{\includegraphics[width=14  cm]{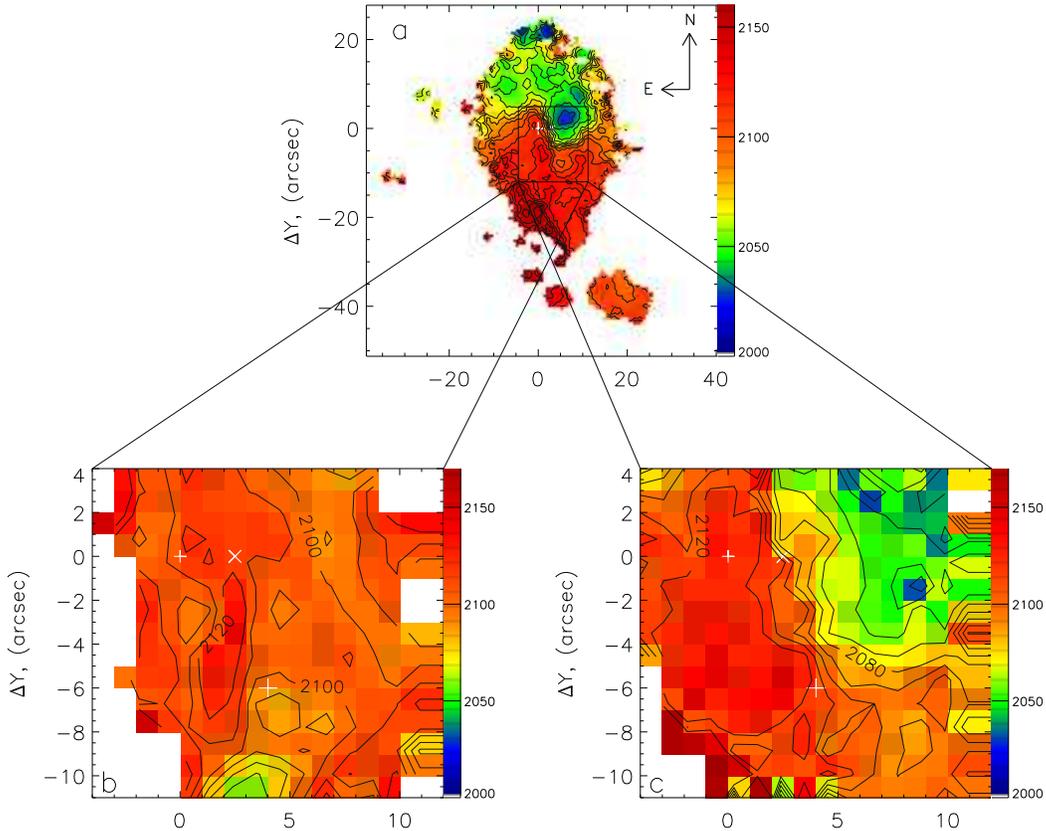}}
 \caption{(a) Large-scale velocity field, as inferred from IFP; (b) and
(c) the velocity field of the stellar and gaseous components for the
central region obtained from MPFS: the pluses indicate the positions
of condensations ``a'' and ``c'' and the cross indicates the position
of the dynamic center.}
\end{figure*}

The observed rotation curve of the galaxy constructed from our data
(crosses) and from the HI data of Taylor et al. (1994) (triangles) is
shown in Fig. 6a. At $R \leq 3$ kpc, the triangle lies well below the
crosses. This is most likely because the 21-cm observations have a
low spatial resolution ($\sim 3 kpc$). At $R \geq 3$ kpc, the
discrepancy between the optical and H I data decreases and does not
exceed $10 \km$. The rotation curve reflects the motions in the inner
disk/ring in the region $0 \leq R \leq 0.8$ kpc and the rotation of
the galactic disk farther out. There is no significant jump in
velocity when passing from one kinematic system to the other, which
most likely suggests that the galaxy's potential is spherically
symmetric on these scales.

We analyzed the observed rotation curve using the models described
by Monnet and Simien (1977). At distances of $0-3$ kpc, it is well
represented by an exponential disk model (the dashed line in Fig.
6a) with the scale factor $h = 0.9$ kpc. This value is close to the
estimates given by Evstigneeva (2000) and Cairos et al. (2001a), if
these are recalculated to our assumed distance to the galaxy. Note
that the photometric profile in the works by Evstigneeva (2000) was
represented by two components: bulge$+$disk. However, our rotation
curve does not confirm the presence of a bulge. If the galaxy
actually had a bulge, then the velocity gradient in the central
region would be much larger than that in our rotation curve.
Therefore, we assume that the increase in brightness compared to the
exponential law observed in the central part of the profiles could be
due to the existence of regions of active star formation. The HI
data show that the rotation curve flattens out somewhere at $8-10$
kpc. Although there may be a small systematic shift between the
optical and radio data, in general, the run of the rotation curve at
large distances from the center ($R \geq 3$ kpc) can be explained by
assuming the presence of an extended spherical isothermal halo. Its
parameters were found to be the following: $r_c = 10$ kpc, $\rho_0 =
0.002\,M_\odot\,\mbox{kpc}^{-3}$. The combined theoretical rotation
curve is indicated in Fig. 6a by the solid line. The overall shape
of the rotation curve is characteristic of late-type galaxies (Amram
and Garrido 2002). The total mass of the galaxy, including the halo,
is $2\cdot10^{10}M_\odot$, and the mass of the disk/ring assuming its
radius to be 0.75 kpc is $\sim 4\cdot 10^8M_\odot$, or 2\% of the
galaxy's mass.

\begin{figure*}
\centerline{\includegraphics[width=14  cm]{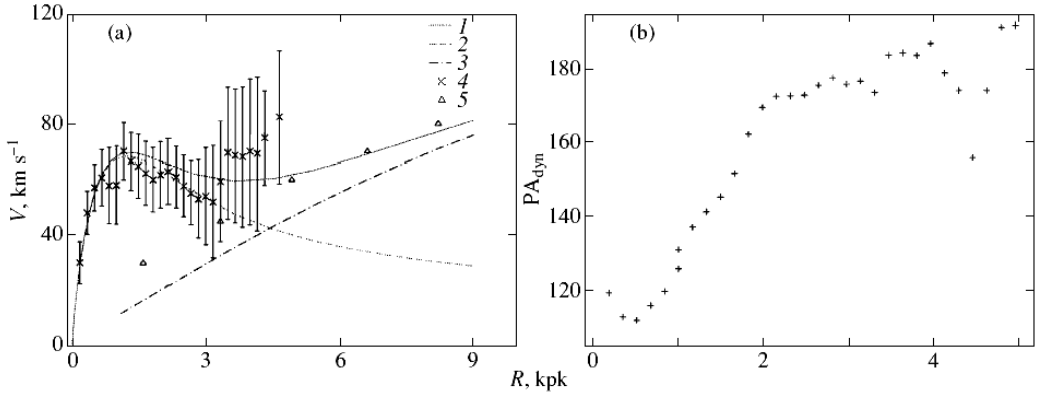}}
 \caption{(a) Rotation curves. 1: the resulting curve; 2: for the disk
component; 3: for the halo; 4: our data; 5: the data from Richter et
al. (1990). (b) The change in the location of the dynamic axis with
radius.}
\end{figure*}

The velocity fields for the stellar and gaseous components in the
central part of the galaxy NGC 7468 are different (Figs. 5b and 5c).
As we see from the figure, the ionized gas in this region rotates
around the major axis, while the stellar velocity field is highly
irregular. The velocity range does not exceed $40 \km$. The observed
velocity variations can be explained by the presence of noncircular
motions. It should also be noted that the contrast of the absorption
features from the old stellar population is low for galaxies of late
morphological types; therefore, the accuracy of the line-of-sight
velocity measurements is lower than that of the usual MPFS
observations of earlier-type galaxies, being $10-20 \km$. We
attempted to describe the stellar velocity field using the method of
``tilted-rings'' with the fixed parameters $PA$ and $i$ corresponding
to those of the galaxy's outer disk. The model describes the
observations satisfactorily, except for the region near condensation
``a'', where large residual velocities ($\sim 50 \km$) are observed.
We found no rotation around the major axis in the stellar velocity
field. Therefore, we conclude that the stars and the gas in the
galaxy's central region belong to different kinematic systems.

\begin{center}
\large \bf DISCUSSION
\end{center}

As was noted in the Introduction, the putative polar ring manifests
itself in the galaxy's images as lobes on its northern and
southern sides. However, analysis of the velocity field that we
constructed from the IFP data and the velocity field derived from
neutral hydrogen (Taylor et al. 1994) shows no peculiar features
near these lobes. The gas in these regions is involved in the
rotation of the galactic disk around the minor axis. Here, there is
no kinematically decoupled subsystem characteristic of polar rings,
which forces us to reject the assumption that NGC 7468 is a galaxy
with an outer polar ring. The nature of the lobes and the
brightening in the south is not quite clear. If we assume, following
Taylor et al. (1993), that the galaxy to the north of NGC 7468 is
spatially close to it (according to this paper, the difference
between their line-of-sight velocities is $73 \km$), then these may
be considered as tidal structures, as may be evidenced by the
deviations of the observed velocities from the circular ones in the
region of the brightening (by $40 \km$) and in the region of the
southern lobe (by approximately $20\km$). On the other hand,
these may be weakly developed spiral arms. Their appearance and the
color bluer than that for the remaining parts of the galaxy (Cairos
et al. 2001b) can serve as evidence for this interpretation.

At the same time, we found a kinematically decoupled gaseous
subsystem, a rotating inner disk, in the central part of NGC 7468
($\sim 1.5$ kpc in diameter). The angle between the galactic disk
and the plane of the inner disk is $86^\circ$. Thus, we can assert
that the galaxy has an inner polar disk (possibly, a ring) rather
than a bar whose presence was used to explain some of the features
in the velocity field constructed from the HI data (Taylor et al.
1994).

According to current views (Bournaud and Combes 2003), a polar
structure in a galaxy can be formed either through accretion from a
neighboring gas-rich galaxy during their close interaction or
through the direct collision of galaxies, which can lead to the
complete absorption of the less massive participant of the collision
by the more massive galaxy. At the observed orientation of the inner
disk, the position of the northern galaxy does not permit us to
consider it a donor galaxy. Thus, we conclude that NGC 7468 collided
with a low-mass gas-rich galaxy that was absorbed. As a result, a
metal-poor polar gaseous disk was formed, and the observed induced
star formation began in the central region of the galaxy (in a ring
with a radius of $\sim 1.2 kpc$). The individual condensations in
this ring are spaced $0.5-1$ kpc apart. Such distances are observed
between neighboring gas-dust clouds. The sizes of the condensations
do not exceed 500 pc, in agreement with the sizes of stellar
associations and stellar complexes. The intensity ratio of the
emission lines suggests the photoionization mechanism of the
emission in these regions.

Note also that the difference between the relative brightnesses of
the condensations in the continuum and in the emission lines may be
due to the age difference between these star-forming regions.
Condensation ``c'', being relatively brighter in the continuum, may
be an older stellar complex at a stage when the surrounding gas had
already been partially swept up to large distances by strong light
pressure light and/or supernova explosions.

\begin{center}
\large \bf CONCLUSIONS
\end{center}

Below, we summarize our results.

(1) Based on the long-slit spectra near \Ha, we found a
kinematically decoupled gaseous subsystem in the central region
rotating around the galaxy's major axis.

(2) Our analysis of the line-of-sight velocity elds obtained using
two-dimensional spectroscopy confirmed the existence of two kinematic
subsystems of ionized gas: one of these is the gaseous disk of the
galaxy, and the other is an inner disk 1.5 kpc in size. The angle
between the planes of the disks is $86^\circ$; i.e., the inner disk
is polar. In the central region of the galaxy, the stars and the
ionized gas belong to different subsystems. (

3) The arc-shaped lobes on the northern and southern sides of
the galaxy are not kinematically decoupled to an extent that the
presence of the polar ring suggested by Whitmore et al. (1990) be
confirmed. They are probably either tidal structures or weakly
developed spiral arms.

(4) The intensity ratio of the forbidden and permitted lines
confirms that the emission in HII regions results from
photoionization. The derived metallicity is lower than the solar
value ($\sim 0.3Z_\odot$).

(5) The detected inner disk allows the galaxy NGC7468 to be
classified as belonging to PRGs. The central collision with a dwarf
galaxy and its capture could be responsible for the formation of the
inner polar disk. The velocity field we derived shows that there is
no reason to believe that the arc-shaped lobes on the northern
and southern sides of the galaxy, which belong to the putative polar
ring (Whitmore et al. 1990), are a kinematically decoupled ring.
These are probably either tidal structures or weakly developed
spiral arms.

\begin{acknowledgements}
 We are grateful to the Large Telescopes Program Committee  (LTPC)
  for allocating observational time on the 6-m
telescope. This study was supported by the Russian Foundation for
Basic Research (project nos. 02-02-16033 and 03-02-06766) and the
Russian Ministry of Education (project no. E02- 11.0-5).
\end{acknowledgements}

{}

\textit{Translated by N. Samus'}

\end{document}